\documentstyle[prl,aps,twoside,multicol]{revtex}

\newcommand{\ba}{\begin{eqnarray}}
\newcommand{\ea}{\end{eqnarray}}
\newcommand{\ra}{\rangle}

\begin{document}

\title{Concatenating Decoherence Free Subspaces with Quantum Error
Correcting Codes}

\author{D.A. Lidar,$^{(1)}$ D. Bacon$^{(1,2)}$ and K.B. Whaley$^{(1)}$}

\address{Chemistry Department$^{(1)}$ and Physics Department$^{(2)}$ \\
The University of California, Berkeley, CA 94720.}

\maketitle

\begin{abstract}
An operator sum representation is derived for a decoherence-free
subspace (DFS) and used to (i) show that DFSs are the class of quantum
error correcting codes (QECCs) with fixed, {\it unitary} recovery
operators, and (ii) find explicit representations for the Kraus
operators of collective decoherence.  We demonstrate how this can be
used to construct a concatenated DFS-QECC code which protects against
collective decoherence perturbed by independent decoherence. The code
yields an error threshold which depends only on the perturbing
independent decoherence rate. \\

\end{abstract}

\begin{multicols}{2}

Decoherence-free subspaces (DFSs) have recently emerged \cite
{Palma:96,Duan:97PRL-Duan:98,Zanardi:97c,Zanardi:97a,Zanardi:98a,Lidar:PRL98}
as an alternative way to protect fragile quantum states against
decoherence, alongside ``conventional'' quantum error correcting codes
(QECCs) \cite {Gottesman:96,Knill:97b} and the new ``dynamical decoupling'' schemes
\cite{Viola:98}. This is of particular importance in quantum
computation, where the promise of a speed-up compared to classical
computers hinges crucially on the possibility to maintain quantum
coherence throughout the computation \cite{Steane:98}. So far, DFSs
and QECCs have been considered as distinct methods, often
characterized as ``passive'' and ``active'' respectively. However, as
we will show here, in fact DFSs can be considered as a special class
of QECCs, characterized as having a particularly simple
form of recovery operators. Conditions for the existence of
non-trivial DFSs are stringent: the decoherence process
should be ``collective'', meaning that the bath couples in a symmetric
way to all qubits.  So far conditions for collective decoherence have
been formulated in a Hamiltonian form \cite{Zanardi:97c,Zanardi:97a},
and in the Lindblad semigroup form
\cite{Zanardi:98a,Lidar:PRL98}. Here we will present an alternative
formulation in terms of the operator sum representation (OSR)
\cite{Kraus:83}, which has the advantage of establishing a direct link
to the theory of QECCs. This OSR formulation enables us to combine
DFSs and QECCs, into a concatenated scheme which can error-correct the
more general physical situation of ``cluster decoherence''. For the
price of longer codewords, this concatenated scheme operates with a
substantially reduced error threshold.

{\it Hamiltonian Formulation of Decoherence Free Subspaces}.---
Conditions for DFSs within the general (non-Markovian) framework of
semigroup dynamics were derived in Ref.~ \cite{Zanardi:97a}. We first
briefly rederive these conditions in a simplified form. Consider a
closed quantum system, composed of a system $S$ of interest defined on
a Hilbert space ${\cal H}$ (e.g., a quantum computer) and a bath
$B$. The evolution of the closed system is given by $\rho
_{SB}(t)={\bf U}\rho _{SB}(0){\bf U}^{\dagger }$, where the
unitary evolution operator (we set $\hbar =1$) is ${\bf U}=\exp
(-i{\bf H}t)$. The full Hamiltonian is ${\bf H}={\bf H}_{S}\otimes
{\bf I} _{B}+{\bf I} _{S}\otimes {\bf H}_{B}+{\bf H}_{I}$, where ${\bf
H}_{S}$, ${\bf H}_{B}$ and ${\bf H}_{I}$ are, respectively, the
system, bath and interaction Hamiltonians, and ${\bf I}$ is the
identity operator. Assuming initial decoupling between system and
bath, the evolution of the closed system is given by: $\rho
_{SB}(t)={\bf {U}}[\rho _{S}(0)\otimes \rho _{B}(0)]{\bf {
U}}^{\dagger }$.  Quite generally, the interaction Hamiltonian can
be written as ${\bf H} _{I}=\sum_{\alpha }{\bf F}_{\alpha }\otimes {\
{\bf B}}_{\alpha }$, where ${\bf F}_{\alpha }$ and ${\bf B} _{\alpha
}$ are, respectively, system and bath operators. Suppose that there
exists a degenerate subset $\{|\tilde{k}\rangle \}$ of eigenvectors of
the ${\bf F}_{\alpha }$'s such that:

\begin{equation}
{\bf F}_{\alpha }|\tilde{k}\rangle =a_{\alpha }|\tilde{k}\rangle
\qquad \forall \alpha , |\tilde{k}\rangle .
\label{eq:DFS-cond}
\end{equation}
If ${\bf H} _{S}$ leaves the Hilbert subspace $\tilde{{\cal
H}}=$Sp$[\{|\tilde{ k}\rangle \}]$ invariant, and if we start within
$\tilde{\cal 
H}$, then the evolution of the system will be {\em decoherence free}
(DF). To show this, expand the initial density matrices of the system
and the bath in their respective bases: $\rho _{S}(0)=\sum_{ij}
s_{ij}|\tilde{i}\rangle \langle \tilde{j}|$ and $\rho
_{B}(0)=\sum_{\mu \nu }{b}_{\mu \nu }|{\mu }\rangle \langle \nu
|$. Using Eq.~(\ref{eq:DFS-cond}), one can write the combined
operation of the bath and interaction Hamiltonians over $\tilde{{\cal
H}}$ as:

\[
{\bf I}_{S}\otimes {\bf H}_{B}+{\bf H}_{I}={\bf I}_{S}\otimes {\bf
H}_{\text{ {\rm c}}}\equiv {\bf I}_{S}\otimes \left[ {\bf
H}_{B}+\sum_{\alpha }a_{\alpha }{\bf B}_{\alpha }\right] .
\]
This clearly commutes with ${\bf H}_S$ over $\tilde{{\cal H}}$.  Thus
since neither ${\bf H}_{S}$ (by our own stipulation) nor the combined
Hamiltonian ${\bf H}_{\rm c}$ takes states out of the subspace:

\begin{equation}
{\bf U}[|\tilde{i}\rangle \otimes |\mu \rangle ]={\bf U}_{S}| \tilde{i}
\rangle \otimes {\bf U}_{{\rm c}}|\mu \rangle ,
\label{eq:UsUc}
\end{equation}
where ${\bf U}_{X}=\exp (-i{\bf H}_{X}t)$, $X = S, c$. Hence it is
clear, given the initially decoupled state of the density matrix, that
the evolution of the closed system will be:
$
\rho_{SB}(t)=\sum_{ij}s_{ij}{\bf U} _{S}|\tilde{i}\rangle \langle
\tilde{j}|{\bf U}^\dagger_{S}\otimes \sum_{\mu \nu }b_{\mu \nu}{\bf
U}_{{\rm c}}|\mu \rangle \langle \nu |{\bf U}^\dagger_{{\rm c}}
.$
It follows using simple algebra that after tracing over the bath:
$
\rho _{S}(t)={\rm Tr}_{B}[\rho _{SB}(t)]={\bf U}_{S}\rho _{S}(0){\bf
U}^\dagger_{S}
$,
i.e., that the system evolves in a completely unitary fashion
on $ \tilde{{\cal H}}$: under the condition of Eq.~(\ref{eq:DFS-cond})
the subspace is DF. As shown in Ref.~ \cite{Zanardi:97a},
Eq.~(\ref{eq:UsUc}) is also a necessary condition for a DFS.

{\it Operator Sum Representation on a Decoherence Free Subspace}.---
In the OSR, the evolution of the density matrix is written as:$\;\rho
_{S}(t)=${\rm Tr}$_{B}[{\bf {U}}(\rho _{S}\otimes \rho _{B}){\bf
{U}}^{\dagger }]=\sum_{a} {\bf {A}}_{a}\,\rho _{S}(0)\,{\bf
{A}}_{a}^{\dagger }$, where the ``{\em Kraus operators}'' are given
by:

\begin{equation}
{\bf A}_{a}=\sqrt{\nu }\langle \mu |{\bf U}|\nu \rangle \;; \qquad
a=(\mu,\nu ) ,
\label{eq:Amunu}
\end{equation}
($|\mu\ra$, $|\nu\ra$ are bath states) and satisfy the normalization
constraint $\sum_{a}{\bf A}_{a}^{\dagger }{\bf A}_{a}={\bf I}_{S}$. 

Let $\tilde{{\cal H}}$ be an $\tilde{N}$-dimensional DFS. In this case
it follows
immediately from Eqs.~(\ref{eq:UsUc}) and (\ref{eq:Amunu}) that the Kraus
operators all have the following representation (in the basis where the first
$\tilde{N}$ states span $\tilde{{\cal H}}$): 

\begin{equation}
{\bf A}_{a}=\left(
\begin{array}{cc}
g_{a}\tilde{\bf U}_{S} & {\bf 0} \\
{\bf 0} & \bar{{\bf A}}_{a}
\end{array}
\right) \,;\qquad g_{a}=\sqrt{\nu }\langle \mu | {\bf U}_{{\rm c}}|\nu
\rangle .
\label{eq:A-block}
\end{equation}
Here $\bar{{\bf A}}_{a}$ is an arbitrary matrix that acts on ${\tilde{{\cal H}}}^\perp$ (${\cal H}={
\tilde{{\cal H}}} \oplus {\tilde{{\cal H}}}^\perp$) and may cause decoherence
there; $\tilde{\bf U}_{S}$ is ${\bf U}_{S}$ restricted to
$\tilde{{\cal H}}$. This simple condition can be summarized as
follows:

{\it Theorem I.} A subspace $\tilde{{\cal H}}$ is a DFS iff all Kraus
operators have an identical unitary representation upon restriction to
it, up to a multiplicative constant.

Thus, in the OSR, the task of identifying a DFS reduces to finding a
subspace in which all the Kraus operators act as the system unitary
evolution operator.  We now give an example for the
important case of collective decoherence (CD). CD is generally
described within the following scenario: the system operators $\{{\bf
F}_{\alpha }\}$ in the interaction Hamiltonian form the Lie algebra
$su(2)$ \cite{Zanardi:97c,Lidar:PRL98}. This means that the
interaction Hamiltonian can be rewritten as:

\begin{equation}
{\bf H}_{I}={\bf S}_{+}\otimes {\bf V}_{+}+{\bf S} _{-}\otimes {\bf
V}_{-}+ {\bf S}_{z}\otimes {\bf V}_{z}.
\label{eq:H_I:CD}
\end{equation}
Here ${\bf S}_{\alpha }=\sum^{K}_{i=1}\sigma _{i}^{\alpha }$ are {\em
global} Pauli spin operators ($i$ is the qubit index) satisfying the
$sl(2)$ commutation relations, and ${\bf V}_{\alpha }$ are the bath
operators coupled to these degrees of freedom. A more restricted case
of CD arises when only phase damping processes are allowed, so that
${\ {\bf V}}_{+}={\bf V}_{-}=0$. We will concentrate on this case, as
it is fully analytically solvable. For simplicity we will assume
throughout that ${\bf H}_{S}=0$.

Pure phase damping on a single qubit is described by the Pauli
$\sigma^z$ matrix. In the standard basis it is easy to verify that the
matrix representation of the global phase damping operator ${\bf
S}_{z}$ is: ${\bf S}_z = {\rm diag}[f(j)]$, where $ f(j)=$ (no. of
0's) - (no. of 1's) in the binary representation of $j$
($j=0...2^{K-1}$).  E.g., for two qubits: ${\bf S}_z = {\rm
diag}[2,0,0,-2]$. Since ${\bf S}_{z}$
is diagonal, the action of the interaction Hamiltonian ${\bf
H}_{I}={\bf S}_{z}\otimes {\bf V}_{z}$ can be written simply as: ${\bf
H}_{I}|j\rangle |\nu \rangle =f(j)|j\rangle \otimes {\bf V} _{z}|\nu
\rangle$. Hence the action of the full Hamiltonian is:
$
{\bf H}|j\rangle |\nu \rangle =|j\rangle \otimes {\bf \Omega }_{j}|\nu
\rangle$ where ${\bf \Omega }_{j}\equiv f(j){\bf V}_{z}+ {\bf H}_{B}
$.
Similarly,
$
{\bf H}^{n}|j\rangle |\nu \rangle =|j\rangle \otimes {\bf
\Omega }_{j}^{n}|\nu \rangle
$,
whence
$
\exp (-i{\bf H} t)|j\rangle
|\nu \rangle =|j\rangle \otimes \exp (-i{\bf \Omega }_{j}t)|\nu
\rangle
$,
so that the Kraus operators can be evaluated explicitly:
$
\langle j^{\prime }|{\bf A}_{a}|j\rangle =\delta _{jj^{\prime
}}\sqrt{\nu } \langle \mu |\exp (-i{\bf \Omega }_{j}t)|\nu
\rangle
$.
Thus, {\em the Kraus operators for pure phase damping CD
have a diagonal matrix representation in the standard basis}, which
can be written compactly as:
$
{\bf A}_{a} ={\rm diag}\left[ g_{a}^{(f_{j})}\right]$,
$g_{a}^{(f_{j})} =\sqrt{\nu }\langle \mu |\exp (- i{\bf
\Omega}_{j}t)|\nu \rangle
$.
For example, in the case of two qubits we obtain
$
{\bf A}_{a}={\rm diag}\left[
g_{a}^{(2)},g_{a}^{(0)},g_{a}^{(0)},g_{a}^{(-2)} \right]
$,
with
$
g_{a}^{(0)}=\sqrt{\nu }\langle \mu |\exp (-i{\bf H}_{B}t)|\nu \rangle
$
and
$
g_{a}^{(\pm 2)}=\sqrt{\nu }\langle \mu |\exp (-i{[{\bf H}}
_{B}\pm 2{\bf V} _{z}]t)|\nu \rangle
$.
By Theorem I the three blocks in ${\bf A}_{a}$ correspond to three
DFSs. Next, consider the 
effect of this OSR on a general density matrix (omitting standard
algebra):
$
\lbrack \rho (0)]_{jk}
\stackrel{t}{\longmapsto }\sum_{a}\left[ {\bf A} _{a}\rho (0){\bf A}
_{a}^{\dagger }\right] _{jk} = [\rho (0)]_{jk}\sum_{\mu \nu }\nu g_{\mu \nu
}^{(f_{j})} {g_{\mu \nu }^{(f_{k})}}^{\ast }
$.
As expected, no mixing of density matrix elements occurs. The time
dependence of each element is determined by the sum in the last
expression, which, motivated by the understanding that decoherence is
taking place, we write formally as a decaying exponential (although
without a Markovian approximation Poincar\'{e} recurrences
may occur). Thus:
$
\exp (-t/\tau _{jk}) \equiv \sum_{\mu \nu }\nu
g_{\mu \nu }^{(f_{j})} {g_{\mu \nu }^{(f_{k})}}^{\ast }
= \sum_{\nu }\nu \langle \nu |\exp \{i t \left[ f(j)-f(k)\right] ({\bf
V}_z + [{\bf V}_z,{\bf H}_B]) + ... \}
|\nu \rangle
$,
with higher order terms, all depending on powers of $f(j)-f(k)$, given
by the Campbell-Hausdorff formula. In agreement with the general theory of DFSs
\cite{Zanardi:97a,Lidar:PRL98}, the DF states are those for which $f(j)=f(k)$, in which case $1/\tau
_{jk}=0$. Normally (i.e., in the ``pointer basis'' \cite{Zurek}), all other states
are expected to have $ 1/\tau 
_{jk}>0$, although in order to verify this one must specify ${\bf
V}_z$. This expectation is confirmed for a harmonic bath
\cite{Palma:96}.

{\it Decoherence Free Subspaces as Quantum Error Correcting Codes}.---
Quantum error correction can be regarded as the theory of reversal of quantum operations on a
subspace \cite{Nielsen:98}. This subspace, ${\cal C}={\rm
Sp}[\{|i_{L}\rangle \}]$, is interpreted as a ``code'' (with codewords
$\{|i_{L}\rangle \}$) which can be used to protect part of the 
system Hilbert space against decoherence (or ``errors'') caused by the
interaction between system and bath. The errors are represented by the Kraus
operators $\{{\bf A}_{a}\}$ \cite {Knill:97b}. To decode the quantum
information after the action of the bath, one introduces ``recovery''
operators $\{{\bf R}_{r}\}$. A QECC is a subspace ${\cal C}$ and a set
of recovery operators $\{{\bf R}_{r}\}$. Ref.~\cite
{Knill:97b} gives two equivalent criteria for the general condition
for QECC. It is possible to correct the errors induced by a given set
of Kraus operators $\{{\bf A}_{a}\}$, (i) iff

\begin{equation}
{\bf R}_{r}{\bf A}_{a}=\left(
\begin{array}{cc}
\lambda _{ra}{\bf I}_{{\cal C}} & {\bf 0} \\ {\bf 0} & {\bf B}_{ra}
\end{array}
\right)  \qquad
\forall r,a , 
\label{eq:RA-block}
\end{equation}
or equivalently, (ii) iff

\begin{equation}
{\bf A}_{a}^{\dagger }{\bf A}_{b}=\left( \begin{array}{cc}
\gamma _{ab}{\bf I}_{{\cal C}} & {\bf 0} \\ {\bf 0} & \bar{{\bf
A}}_{a}^{\dagger }\bar{{\bf A}}_{b} \end{array}
\right) \qquad
\forall a,b .
\label{eq:cond2}
\end{equation}
In both conditions the first block acts on ${\cal C}$; ${\bf B}_{ra}$
and $ \bar{{\bf A}}_{a}$ are arbitrary matrices acting on ${\cal
C}^{\perp }$ ($ {\cal H}={\cal C}\oplus {\cal C}^{\perp }$). Let us
now explore the relation between DFSs and QECCs. First of all, it is
immediate that DFSs are indeed a valid QECC. For, given the (DFS-)
representation of ${\bf A}_{a}$ as in Eq.~(\ref{eq:A-block}), it
follows that Eq.~(\ref{eq:cond2}) is satisfied with $ \gamma
_{ab}=g_{a}^{\ast }g_{b}$. Note, however, that unlike the general QECC
case which has a full-rank matrix $\gamma _{ab}$, in the DFS case this
matrix has rank 1 (since the $a^{\rm th}$ row equals row 1 upon
multiplication by $g_{1}^{\ast }/g_{a}^{\ast }$), implying that a DFS
is a highly {\em degenerate} QECC \cite{Gottesman:96,Knill:97b}.

A DFS is an unusual QECC in another way: decoherence does not affect
a perfect DFS {\em at all}. Since they are based on a perturbative
treatment, other QECCs (e.g., stabilizer or GF(4) codes
\cite{Gottesman:96}) are specifically constructed to improve the fidelity to a
given order in the error rate, which therefore always allows for some
residual decoherence to
take place. The absence of decoherence to any order for a perfect DFS is
due to the existence of symmetries in the system-bath
coupling which allow for an {\em exact} treatment. These symmetries are
ignored by perturbative QECCs either for the sake of generality, or because
they simply do not exist, as in the case of independent couplings. Given a
DFS, the only ``errors'' that can take place involve the unitary rotations
of codewords (basis states $\{| \tilde{i}\rangle \}$ of $\tilde{{\cal
H}}$) inside the DFS, due to the system Hamiltonian ${\bf H}_{S}$
(this may actually be the desired evolution if one is implementing a
computation inside $\tilde{{\cal H}}$ using ${\bf H}_{S}$). Thus, the
complete characterization of DFSs as a QECC is given by the 
following:

{\it Theorem II}. Let ${\cal C}$ be a QECC for error operators $\{{\bf
A} _{a}\}$, with recovery operators $\{{\bf R}_{r}\}$. Then ${\cal C}$
is a DFS iff upon restriction to ${\cal C}$, ${\bf R}_{r}\propto \tilde{\bf
U}_{S}^{\dagger }$ for all $r$.

{\it Proof}. First suppose ${\cal C}$ is a DFS. Then by Eqs.~(\ref
{eq:A-block}) and (\ref{eq:RA-block}), 
\[
{\bf R}_{r}\left(
\begin{array}{cc}
g_{a}\tilde{\bf U}_{S} & {\bf 0} \\
{\bf 0} & \bar{\bf A}_{a}
\end{array}
\right) =\left(
\begin{array}{cc}
\lambda _{ra}{\bf I}_{C} & {\bf 0} \\
{\bf 0} & {\bf B}_{ra}
\end{array}
\right) .
\]
To satisfy this equation, it must be true that \[
{\bf R}_{r}=\left(
\begin{array}{cc}
\frac{\lambda _{ra}}{g_{a}}\tilde{\bf U}_{S}^{\dagger } & {\bf C}_{r} \\ {\bf
D}_{r} & {\bf E}_{r}
\end{array}
\right) .
\]
The condition $g_{a}\tilde{\bf U}_{S}^{\dagger }{\bf D}_{r} = {\bf 0}$
implies ${\bf D}_{r}={\bf 0}$ by unitarity of $\tilde{\bf
U}_{S}$. Also, since $\bar{\bf A}_{a}$ is arbitrary, generically the
condition ${\bf C}_{r}\bar{\bf A}_{a}={\bf 0}$ implies ${\bf
C}_{r}={\bf 0}$. Thus upon restriction to ${\cal C}=\tilde{{\cal H}}$,
indeed ${\bf R}_{r} \propto \tilde{\bf U}^\dagger_{S}$ (by unitarity
of $\tilde{\bf U}_{S}$, $|\lambda _{ra}/g_{a}|=1$). Now suppose ${\bf
R}_{r}\propto \tilde{\bf 
U}^\dagger_{S}$. The very same argument applied to ${\bf A}_{a}$ in
Eq.~(\ref{eq:RA-block}) yields ${\bf A}_{a}\propto \tilde{\bf U}_{S}$ upon
restriction to ${\cal C}$. Since this is exactly the condition
defining a DFS in Eq.~(\ref{eq:A-block}), the theorem is proved.

We conclude that DFSs are a particularly simple instance of general
QECCs, where upon restriction to the code subspace, all recovery
operators are proportional to the inverse of the system evolution operator.

{\it Quantum Error Correction On A Decoherence Free Subspace}.---
A non-ideal DFS will still be subject to some decoherence
\cite{Lidar:PRL98}.  DFSs are efficient under conditions in which each
qubit couples to the same environment (collective
decoherence). Ordinary QECCs are designed to be efficient when each
individual qubit couples to a different environment (independent
decoherence). While neither code is efficient in the extreme limit when the
other is, QECCs will still work for correlated errors
\cite{Gottesman:96,Knill:97b}, whereas DFSs 
will not work in the independent error case \cite{Lidar:PRL98}. One
would generally 
expect the likelihood of $K$ qubits 
collectively coupling to the same environment to decrease with
increasing $K$ \cite{Palma:96}. Thus an interesting situation
(``cluster decoherence'') arises when small blocks of qubits undergo
collective decoherence (e.g., groups of neighboring identical atoms on
a polymer chain), while this symmetry is broken perturbatively by
independent decoherence between blocks. Here we show how by adding an
additional layer of QECC encoding, the DFS can be stabilized against
such computational errors.

In the collective decoherence case [Eq.~(\ref{eq:H_I:CD})], the
smallest DFS which can encode one logical qubit is made up of four
physical qubits \cite{Zanardi:97c}. Consider an operator basis which
covers all possible errors which can occur on this 4-qubit DFS. From
the discussion above, we know that a DFS has Kraus operators which all
are direct sums of a fixed unitary transformation on $\tilde{{\cal
H}}$ and variable transformations (whose exact form is irrelevant) on
$\tilde{{\cal H}}^{\perp }$. Thus, perturbative errors (of size
$\epsilon $) on a DFS can conveniently be represented by Kraus
operators with the following structure:

\begin{equation}
{\bf A}_{a}=\tilde{{\bf A}}_{a}+\epsilon \left( \begin{array}{cc}
{\bf Q}_{1} & {\bf Q}_{2} \\
{\bf Q}_{3} & {\bf Q}_{4}
\end{array}
\right) .
\label{eq:A-pert}
\end{equation}
Here $\tilde{{\bf A}}_{a}$ represents the (dominant) contribution due
to ideal collective decoherence, and the second matrix represents the
symmetry breaking perturbation. ${\bf Q}_{1}$ (a 2 $\times $ 2 matrix
for the 4-qubit case) acts just on the DFS; ${\bf Q}_{2}$ (2 $ \times
$ 14) takes states from $\tilde{{\cal H}}^{\perp}$ into $\tilde{ {\cal
H}}$, but we need not worry about this as a separate process, since it
can also be corrected using QECC inside $\tilde{{\cal H}}$; ${\bf Q}_{3}$
(14 $\times $ 2) takes states from $\tilde{{\cal H}}$ into
$\tilde{{\cal H}}^{\perp }$; ${\bf Q}_{4}$ acts just on $\tilde{{\cal
H}}^\perp$ and is irrelevant to our discussion. Thus, to first order in time, all of the relevant errors can
be enumerated as: (i) Independent errors acting on the encoded DFS
states. A basis for these errors are the Pauli operators: ${\bf X}$,
${\bf Z}$, ${\bf Y }={\bf XZ}$, and ${\bf I}$ acting {\em only} on the
DFS qubits $ |0_{L}\rangle $ and $|1_{L}\rangle $ (e.g., ${\bf
X}|0_{L}\rangle =|1_{L}\rangle $). (ii) Errors which take the system
into $\tilde{{\cal H}} ^{\perp }={\rm
Sp}[\{|j_{L}\rangle\}_{j=2}^{15}]$. Define 14 operators ${\bf 
P}_{j}=|j_{L}\rangle (\langle 0_{L}|+\langle 1_{L}|)$, ${\bf 
P}_{j}: \tilde{{\cal H}} \mapsto \tilde{{\cal H}} ^{\perp }$, out of a
total of 28 in the $ {\bf Q}_{3}$ block. In order to cover all possible
errors which might take the DF states outside of $\tilde{\cal H}$, it
suffices to consider the effect of ${\bf P}_{j}$ and
${\bf P}_{j}{\bf Z}$, where ${\bf Z }$ is the phase error operator
acting on $\tilde{\cal H}$ as defined above. The possible errors which
can occur on our 4-qubit DFS are thus given by:
${\cal E} = \{{\bf X},{\bf Y},{\bf Z},{\bf P}_{j},{\bf P}_{j}{\bf
Z}\}$. The task is now to find an appropriate error correction scheme. To
do so, we may use the DF states $|0_{L}\rangle $
and $|1_{L}\rangle $ to construct the well-known ``perfect'' $5$-qubit
QECC \cite{Laflamme:96}. This concatenation yields new encoded states
$|0_{E}\rangle $ and $ |1_{E}\rangle $, composed of 20 physical
qubits. In the standard approach to correcting errors with
the $5$ qubit code, one uses a quantum network to calculate the
syndrome on ancilla qubits, and uses this syndrome to apply
the appropriate correction procedure on the encoded qubits. Thus, in
order to apply a similar procedure to correct for the standard bit
flip errors which now act on our 4-qubit DF states, we simply convert
the gates used in the standard error correction procedure to gates
which perform the same operations on $|0_{L}\rangle $ and $
|1_{L}\rangle $ and which do not disturb the states in $\tilde{ {\cal
H}}^{\perp }$. Under these conditions, it is obvious that ${\bf X}$,
${\bf Y}$, ${\bf Z}$ can be corrected if the DF qubits decohere
independently (the condition on standard QECC), as we stipulated.
Furthermore, consider the following modified controlled-not gate:
${\bf C}|0_{L},0_{L}\rangle =|0_{L},0_{L}\rangle $,
${\bf C} |1_{L},0_{L}\rangle =|1_{L},0_{L}\rangle $,
${\bf C}|j_{L},0_{L}\rangle =|j_{L},j_{L}\rangle $,
plus unspecified operations (which do not have to concern us) which ensure
that $ {\bf C}$ is unitary. Using this gate, one can correct for the
effect of ${\bf P}_{j}$ on the system as follows: attach an ancilla DF qubit in
the $|0_{L}\rangle $ state, perform ${\bf C}$ on this state, and
repeat the procedure on every DF qubit of the 5-qubit codeword.  This detects whether one of
the ${\bf P}_{j}$ errors has occurred. Essential to this procedure is
the fact that we do not disturb the original $ |0_{L}\rangle $ and
$|1_{L}\rangle $ states. Once an error of this type has been detected,
one may recover the erred state to $|0_{L}\rangle$. The error has then
been reduced to a 
standard Pauli one on a DFS state, and can be fixed by QECC. For
example (the first qubit, $x=0$ or 1, belongs to the
codeword and the second is the ancilla):
$|x_L\ra |0_L\ra \stackrel{{\bf P}_{2}}{\mapsto}
(|x_L\ra + |2_L\ra ) |0_L\ra \stackrel{{\bf C}}{\mapsto}
|x_L\ra |0_L\ra + |2_L\ra |2_L\ra
$.
Next the ancilla is measured: if the result is 0, no error has
occured; if it is 2 then the state is recovered to $|0_L\ra$ and the
resulting standard Pauli error can be fixed by comparison of the erred
DF qubit to the other 4 DF qubits in the codeword.
Finally,
the extreme case of single independent physical qubit errors is also
dealt with by the concatenated DFS-QECC code, because the above
procedure automatically also corrects the errors which are the subject
of the standard QECC procedures, representing them as linear
combinations drawn from our natural DFS basis of errors ${\cal E}$. Of
course, the 20-qubit code presented above is less efficient than the
standard 5-qubit QECC code if these independent errors are the dominant ones.

The final question regarding the concatenated DFS-QECC scheme concerns
the threshold for fault tolerant quantum computation
\cite{Preskill:97a}.  The threshold probability of error has
been estimated to be on the order of $10^{-6}$ per operation
\cite{Knill:98a}. In the present context it is simplest to discuss
this issue in the language of the Markovian semigroup master
equation. Following the notation of Ref.~\cite {Lidar:PRL98}, we
consider error generators $\{{\bf F}_{\alpha }\}$ yielding a
decohering term $[{\bf F}_{\alpha },\rho (t){\bf F}_{\beta }^{\dagger
}]$ in the master equation. Assume these generators produce
errors with a rate $\lambda $. Then the fidelity $F(t)={\rm Tr}[\rho
_{S}(0)\rho _{S}(t)]$ is generally reduced by a term of $O(\lambda
^{2}t)$.  Upon inclusion of a symmetry-breaking perturbation by error
generators $\{{\bf G }_{p}\}$ of order $\epsilon $, with $\lambda \gg
\epsilon $, we find, following the arguments in
Ref.~\cite{Lidar:PRL98}, that this leads to a decrease in the
fidelity, not of order $O(\lambda \epsilon t)$ as one might naively
expect (since $[{\bf G}_{p},\rho (t){\bf F}_{\beta }^{\dagger }]$
terms have appeared), but, remarkably, only of order $O(\epsilon
^{2}t)$ (due to the $[{\bf G}_{p},\rho (t){\bf G}_{q}^{\dagger }]$
terms) \cite{comment}.

Perturbing a DFS thus produces error rates which are solely
proportional to the strength of the additional perturbing process, and
which are not dependent on the rate of the error process which
generates the DFS. Applying a QECC to this system, as
discussed above, improves the fidelity to $1-O(\epsilon
^{4}t^{2})$. Now in quantum computation one generally envisions a
realization of a quantum computer which has an {\it a priori} low
decoherence rate $\lambda$. But in the ordinary QECC case, it is this
rate $\lambda $ which sets the error threshold. Hence, in the scenario
envisioned here, we have effectively decoupled the rate $\lambda $
from the error threshold, and a significant improvement may therefore
result, if the additional perturbation $\epsilon $ is sufficiently
small. Concatenation of a DFS with QECC thus has the potential to
achieve the goal of truly fault tolerant quantum computation, not just
quantum memory.

Acknowledgements.---
This material is based upon work supported by the U.S. Army
Research Office under contract/grant number DAAG55-98-1-0371, and in part by
NSF CHE-9616615. We thank Dr. P. Zanardi for helpful
correspondence.

\end{multicols}

\end{document}